\documentclass[preprintnumbers,amsmath,amssymb,unsortedaddress,showpacs]{revtex4}
\usepackage{graphicx,color}
\usepackage{amsmath,amssymb,latexsym}

\begin{document}

\title{Active Brownian Particles Escaping a Channel in Single File}
\author{Emanuele Locatelli}
\email{emanuele.locatelli@pd.infn.it}
\affiliation{
Dipartimento di Fisica e Astronomia `G. Galilei' and Sezione CNISM,
Universit\`a di Padova,
Via Marzolo 8, I-35131 Padova, Italy
}

\author{Fulvio Baldovin}
\email{baldovin@pd.infn.it}
\affiliation{
Dipartimento di Fisica e Astronomia `G. Galilei', Sezione INFN, and Sezione CNISM,
Universit\`a di Padova,
Via Marzolo 8, I-35131 Padova, Italy
}

\author{Enzo Orlandini}
\email{orlandini@pd.infn.it}
\affiliation{
Dipartimento di Fisica e Astronomia `G. Galilei', Sezione INFN, and Sezione CNISM,
Universit\`a di Padova,
Via Marzolo 8, I-35131 Padova, Italy
}

\author{Matteo Pierno}
\email{matteo.pierno@unipd.it}
\affiliation{
Dipartimento di Fisica e Astronomia `G. Galilei' and Sezione CNISM, 
Universit\`a di Padova,
Via Marzolo 8, I-35131 Padova, Italy
}

\date{\today}

\begin{abstract}
Active particles may happen
to be confined in channels so narrow that they cannot overtake each
other (Single File conditions). 
This interesting situation reveals nontrivial
physical features as a consequence of the strong inter-particle
correlations developed in collective rearrangements. 
We consider a minimal model for active Brownian particles with the aim
of studying the modifications introduced by activity with respect to the classical
(passive) Single File picture. 
Depending on whether their motion is
dominated by translational or rotational diffusion, 
we find that active Brownian particles in Single File 
may arrange into clusters which are 
continuously merging and splitting ({\it active clusters})
or merely reproduce passive-motion paradigms, respectively.
We show that activity convey to self-propelled
particles a strategic advantage for trespassing narrow channels
against external biases (e.g., the gravitational field). 
\end{abstract}

\pacs{64.75.Gh, 47.63.Gd, 47.57.eb}


\maketitle

\section{Introduction}

The study of the physical properties of living matter has received
great attention in the last decade. Beside the relevance in Biology,
Physiology and Medicine, these systems can show intriguing new
Physics. In particular, active or self-propelled particles (SPP) --
individual units that can transform chemical energy into kinetic
energy -- exhibit a wealth of non-equilibrium phenomena, like collective motion~\cite{Vicsek}, dynamical patterns
formation~\cite{Cates_et_al_2010, Berthier_13, Yeomans_12}, 
and unusual rheological properties~\cite{Hatwalne}. Examples of SPP are microbes and bacteria~\cite{Berg2004}, algae~\cite{Poli_et_al_2009}, molecular motors~\cite{Surrey}, but also artificial
colloidal microswimmers can be designed to achieve reliable and
reproducible examples of self-propulsion~\cite{Paxton_et_al_2004}.  
In physically- and biologically-relevant situations~\cite{Mannik,Biondi,Lebleu,Binz,Golestanian_Soto_2013,Costanzo1,Costanzo2}, SPP can be constrained to move in channels so narrow that the mutual passage of the constituents is prohibited. Whenever this happens we say that they form an \emph{Active Single File} (ASF) system. It is known that extreme confinement conditions in passive systems generate strong inter-particle correlations, leading to subdiffusion~\cite{Harris,Burada,Wei,Kollmann_2003}.  
Under strong confinement as well as biases imposed by external fields or environmental conditions, the enhanced mobility characteristic of active particles may lead to new collective dynamical properties.
These, in turns, may strongly affect the the system's transport behavior, e.g., in channel-emptying processes, or the microbiology of soil during infections or for filtration and purification
applications~\cite{Mannik}.

Although hydrodynamic effects are expected to play a role under
severe geometric constraints \cite{Drescher1,Zottl}, in favor of simplicity and
analytical tractability in the present context we consider a
minimal Brownian model for active motion \cite{tenHagen}.
Indeed, a major goal in this work is to study how Single File
properties are modified when self-propelled particles are present. We
tackle this issue by investigating
the interplay between the increased mobility due to activity and the
hindering effects coming from the Single File condition.   
In particular, with both open channel ends, we concentrate on the
dependence of its emptying from either the degree of activity or the 
rotational vs translational character of the active motion. 
We find that the active particles
undergo a dynamical transition from rotational to translational behavior, which also
corresponds to a structural change for the ASF
system, with the formation of ``active clusters'' continuously merging and splitting.  
Albeit not significantly improved by the existence of clusters, 
translational active particles expose a significant capability of trespassing channels 
against external biases.

\section{A theoretical model for active motion}
\label{sec:model}
On the basis of Ref.~\cite{tenHagen}, 
we consider a minimal model
where 2-dimensional overdamped active particles in contact with a thermal bath
are self-propelled by a constant force along a direction specified by
an angular coordinate.
An active particle of radius $R$ is
viewed as characterized by a unit vector 
$\mathbf{u}(t)\equiv(\cos\theta(t),\sin\theta(t))$ in the $x y$ plane,
defining the direction of the active force of intensity $F_a$.
The particle is subjected to both translational and rotational
diffusion, with coefficients $D_t$ and $D_r$, respectively.
Once projected along the channel direction, which we assume to
coincide with the $x$-axis, 
the equations of motion become
\begin{equation}
  \left\{ \begin{array}{l}
    \dot{x}(t)=\dfrac{D_t}{k_B T}\left[
      F_{a}\cos\theta(t)
      +F_{e}
      \right]
    +\xi_t(t),
    \\
    \\
    \dot{\theta}(t)=\xi_r(t),
  \end{array} \right.
\end{equation}
where $k_B$ is the Boltzmann constant, $T$ the heat-bath temperature, $F_e$ is the external uniform force, and 
both $\xi_t(t)$ and $\xi_r(t)$ are Gaussian white noise,
$\langle\xi_t(t)\,\xi_t(t^\prime)\rangle={2D_t}\,\delta(t-t^\prime)$,
$\langle\xi_r(t)\,\xi_r(t^\prime)\rangle={2D_r}\,\delta(t-t^\prime)$,
representing a random force and a random torque, respectively.
In particular, $D_t$ is assumed to satisfy the Einstein relation 
$\zeta\,D_t=k_B T$  where $\zeta$ is the translational friction coefficient that, 
for a spherical particle of radius $R$ in a fluid with viscosity $\eta$ is equal to $6\pi R \eta$.
The active force is related to
the modulus of the drift velocity of the active
particle $v_a$, $F_a=\zeta\,v_a$, in the absence of external fields.

\subsection{Linear vs rotational diffusion: identification of a
  crossover line}

As customary  we quantify the particle's activity
through the P\'eclet number, namely the ratio between advection and diffusion \cite{Bruus}:
\begin{equation}
\mbox{Pe} \equiv \frac{v_{a}R}{D_{t}} = \frac{F_a R}{k_B T} \geq0.
\end{equation}
Within our model, Pe can also be interpreted as the ratio between the work done by
the active force over a distance of the order of the particle's size
and the thermal energy. 
At variance with Ref. \cite{tenHagen}, 
we do not fix the ratio between
$D_r$ and $D_t$. Instead, we encode it into a second adimensional
parameter, 
\begin{equation}
\mbox{Ro} \equiv R^2 D_r/D_t\geq0,
\end{equation}
which we name the {\it rotation number} and that expresses the predominance of the
rotational over the translational diffusion. 
Equivalently, Ro may be seen as the ratio between the translational
and the rotational time scales, $\tau_t=R^2/2D_t$ and $\tau_r=1/2D_r$, respectively. 
${\rm Pe}$ and ${\rm Ro}$ can be easily related to other quantities, 
like, e.g., the persistence length $L_p\equiv v_a/D_r$ which is often used in simulation
as a control parameter: Namely,
\begin{equation}
{L_p}=\frac{\rm Pe}{\rm Ro}\;R.
\end{equation}
The main advantage in using Pe and Ro is that we have two independent
control parameters which neatly separate the effects of the activity and
of the rotational diffusion on the particle dynamics.
In a run-and-tumble model  
in which the rotational dynamics is typically defined through a shot noise, 
the rotation number would be defined as ${\rm Ro}\equiv R^2\alpha/D_t$, 
where $\alpha$ is the tumbling rate \cite{Tailleur_2008,cates_2012}.

Once averaged over a uniform distribution of the initial angles
$\theta(0)=\theta_0$, there is no preferred direction for the active
motion, and the mean square displacement from the initial
position $x(0)=x_0$ evolves as
\begin{eqnarray}
\langle(x(t)-x_0)^2\rangle&=&
2D_t t
+\frac{\mbox{Pe}^2}{\mbox{Ro}^2} \; R^2
\left[D_rt+e^{-D_rt}-1\right]\\
&\simeq &2D_at,
\end{eqnarray}
where the last approximation is valid for $t\gg\tau_r$, and we have
defined an effective, active diffusion coefficient as
\begin{equation}
D_a\equiv D_t \left(1+\frac{\mbox{Pe}^2}{2\,\mbox{Ro}} \right).
\end{equation}
Besides $\tau_t$ and $\tau_r$, we thus have also an active time scale
$\tau_a=R^2/2D_a$.  Notice that as a consequence of the random
direction of the active force, instead of being related to the drift,
$\tau_a$ is associated to an active diffusion process.

\begin{figure}
\centering
\includegraphics[width=0.75\columnwidth]{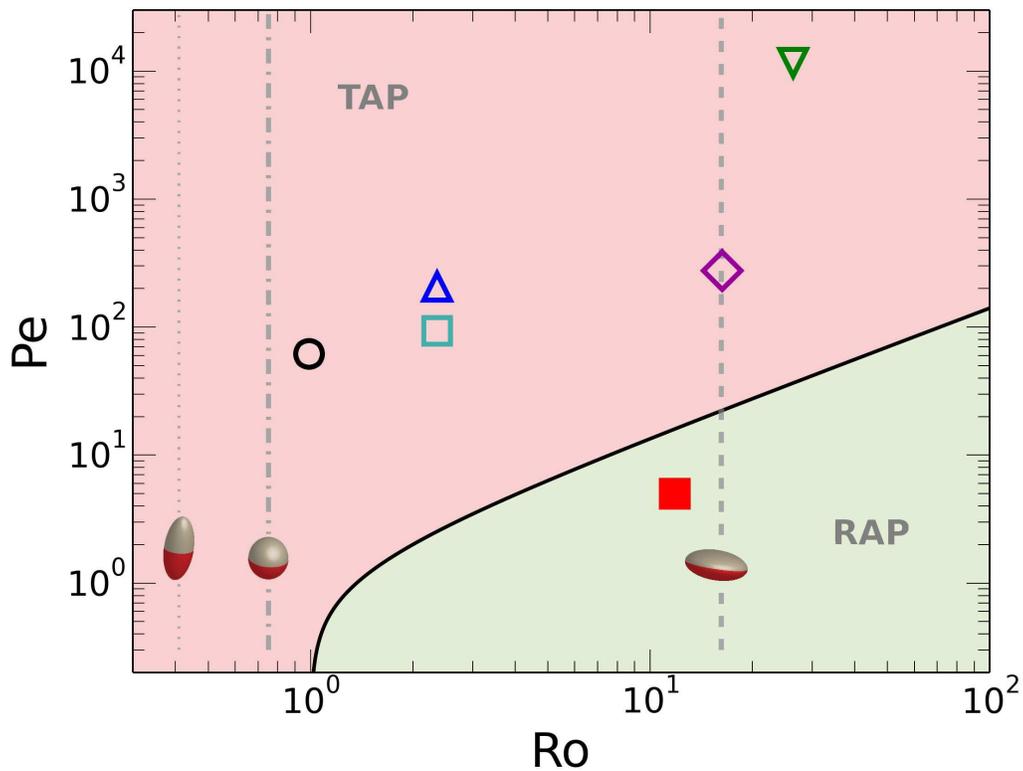}\\
\caption{(Color online) Ro-Pe phase diagram for SPP. 
The  full line, given by Eq.~(\ref{eq:transition_line}) with ${\rm Ro}>1$, distinguishes TAP's from  RAP's.
  Symbols refer to different microorganisms. 
  Circle is \emph{V. Alginolyticus}~\cite{Xie}; Upward triangle is
  \emph{P. Putida}~\cite{Duffy}; Empty square is wild type
  \emph{E. Coli}; Filled square is RAP mutant
  \emph{E. Coli}~\cite{Saragosti}; Diamond is
  \emph{R. Spheroides}~\cite{Rosser}; Downward triangle is
  \emph{C. Reinhardtii}~\cite{Drescher}. Vertical lines represent Janus
  microswimmers: Prolate Spheroids (dotted); Spheres
  (dash-dotted); Oblate Spheroids (dashed).
}
\label{fig:elastic}
\end{figure}

The influence of the
reorientational process on the translational dynamics depends on the
typical timescale of the rotational diffusion as well as on the
magnitude of the active force. It is thus defined  by the ratio between
the typical timescales of rotational and translational motion,
$\tau_{r}$ and $\tau_{a}$ respectively. 
For $\tau_a\ll\tau_r$ the motion is dominated by (active) translational diffusion
with a predominance of straight trajectories over bends;
Vice versa, as  $\tau_a\gg\tau_r$ rotational diffusion become prevalent.
The condition
$\tau_{r}/\tau_{a} = 1 $ leads to a relation between P\'eclet and
Rotation numbers
\begin{equation}
\mbox{Pe}
=\left\{
\begin{array}{ll}
\textrm{Ro}\;\sqrt{2\left(1-\frac{1}{\textrm{Ro}}\right)}
& \textrm{if } \textrm{Ro}>1\\
0
& \textrm{if } 0\leq\textrm{Ro}\leq1
\end{array}
\right.
\label{eq:transition_line}
\end{equation}
that is graphically reported in Fig.~\ref{fig:elastic}.
This relation divides the plane Ro-Pe in two distinct regions and suggests a way to 
classify the active particles of our model as follows:
A Self-Propelled particle may be called {\it Translational Active Particle -- TAP} ({\it Rotational Active Particle -- RAP}) 
if it lies above (below) the line. Note that compared to their characteristic translational timescale
TAP's change the direction of the active force slowly,
hence they 
exhibit long periods of straight motion;
On the other hand rotational diffusion is predominant for RAP's dynamics.
Although in our case the particle orientation is not abruptly ``tumbled'' through a shot noise term, 
we may thus associate TAP's to {\it runners} and RAP's to {\it tumblers} 
with respect to run-and-tumble models \cite{cates_2012}.  \\
Available data for bacteria and
algae indicate that these microorganisms have a tendency to perform long
period of straight motion. Consistently, in the Ro-Pe representation
they are classified as TAP's (See empty symbols in
Fig.~\ref{fig:elastic}). 
On the other hand, bacteria mutants obtained through genetic
engineering may
become RAP's (See filled square in Fig.~\ref{fig:elastic}). 
For artificial microswimmers (Vertical lines in Fig.~\ref{fig:elastic}), Ro is
fixed and depends on the shape, while Pe can be tuned by varying, for
example, the ${\rm H_{2}O_{2}}$ concentration~\cite{Howse_et_al_2007}. 
Eq.~(\ref{eq:transition_line}) 
is particularly interesting when 
studying an ASF system. Indeed,
in passing from RAP's to TAP's
it identifies a crossover line which singles out the formation of
dynamical aggregates.

\section{Active Brownian particles escaping a channel in Single File}

In analogy with passive systems one would expect that interacting Self
Propelled particles under Single File condition (ideal representation
of an extreme channel-like confinement) display significant changes
with respect to the dynamical properties in the bulk. By focusing in
particular on the emptying process of the Single File we show that the
distinction between RAP's and TAP's determines
the main properties of the system such as
dynamical clustering and the mean emptying time either in the presence or
in the absence of an external bias.

\subsection{Analytical derivation of the channel's emptying probability}
\label{sec:analytical}
To properly address the problem of the emptying process of an active
system under Single File constraint it is convenient to look first
at the corresponding problem for a system of passive particles.  
Single File conditions force particles to maintain
spatial order thus generating a complex correlation structures in the
system. As a result, particles perform a subdiffusive motion
which lasts for a very long time if their number is large
enough~\cite{Delfau}. In spite of this, passive Single File evolution
can be mapped onto normal diffusion thanks to the Reflection
Principle. It has been Jepsen~\cite{Jepsen} the first to point out
that it is
possible to map the Single File diffusion onto a free particles
diffusion in which at each collision particles simply exchange their
labels. This is equivalent to impose reflecting boundary conditions at
the planes $x_i(t)=x_j(t)$ at which particles $i$ and $j$
collide~\cite{Rodenbeck}. 

The combinations of all possible reflections are given by the
permutations of the coordinates, so that the probability 
$P_N(\mathbf x,t|\mathbf x^0,N,L)$ 
of finding the Single File array in positions $\mathbf x\equiv(x_1,x_2,\ldots,x_N)$ at time
$t$ within a channel centered in $x=0$ of width $L$, 
given that the initial coordinates a time zero are 
$\mathbf x^0\equiv(x_1^0,x_2^0,\ldots,x_N^0)$ 
with $-L/2\leq x_1^0<x_2^0<\ldots<x_N^0\leq L/2$, 
is provided by the expression \cite{Rodenbeck}
\begin{equation}
\label{eq_joint}
P_N(\mathbf x,t|\mathbf x^0,N,L)=
  \left\{ \begin{array}{l}
    \displaystyle
    \sum_{\pi\in \Pi_N}
    \;\prod_{k=1}^N
    P_1(x_{\pi(k)},t|x_k^0,1,L)
    \quad\textrm{if }
    \mathbf x\in \mathcal D,
    \\
    \\
    0
    \qquad \qquad \qquad \qquad \qquad \;\;\;\;\;\;\; \textrm{if }
    \mathbf x\notin \mathcal D,
  \end{array} \right.
\end{equation}
where $\Pi_N$ is the set of permutations of $N$ objects, $\mathcal D$ represents
the set of allowed configurations, 
$\mathcal D\equiv\{\mathbf x\in\mathbb R^N:
-L/2\leq x_1<x_2<\ldots<x_N\leq L/2\}$, 
and $P_1(x_k,t|x_k^0,1,L)$ is the Green function for the
free single-particle problem with absorbing boundaries at $x=\pm L/2$.
Although exact, Eq. (\ref{eq_joint}) is hard to handle both
analytically and numerically because of the restriction on the
allowed configurations. 
However, if one is interested in the survival
probability of {\it all} the $N$ particles within the channel,
\begin{equation}
S_N(t|\mathbf x^0,N,L)
=
\int_{\mathcal D}
d\mathbf x
\;P_N(\mathbf x,t|\mathbf x^0,N,L),
\end{equation}
permutations in Eq. (\ref{eq_joint}) allow to remove the particles
order constraint and one obtains
\begin{equation}
\label{eq_surv_N}
S_N(t|\mathbf x^0,N,L)
=
\prod_{k=1}^N
S_1(t|x_k^0,1,L),
\end{equation}
where $S_1(t|x_k^0,1,L)$ is the survival probability of a single free particle
originally posed at $x_k^0$ and absorbed in $\pm L/2$.

For the study of the emptying process of a Single File system of identical
particles 
Eq. (\ref{eq_surv_N}) turns out to be a fundamental simplification. 
If we indicate with $S_n(t|\mathbf x^0,N,L)$ the probability of having
{\it at least} $0\leq n\leq N$ particles within the channel at time
$t$, given that initially $N$ particles are posed in position
$\mathbf x^0\equiv(x_1^0,x_2^0,\ldots,x_N^0)$ within a channel of length $L$, 
the channel emptying probability becomes 
\begin{equation}
1-S_1(t|\mathbf x^0,N,L).
\end{equation}
Now, $S_n(t|\mathbf x^0,N,L)$ can be viewed as the sum
\begin{equation}
\label{eq_s_n}
S_n(t|\mathbf x^0,N,L)
=\sum_{k=n}^Np_k(t|\mathbf x^0,N,L),
\end{equation}
where $p_n(t|\mathbf x^0,N,L)$ is the probability of having
$n$ particles within the channel at time $t$, given that at time zero
the $N$ particles were located at $\mathbf x^0$.
Another way of defining $p_n(t|\mathbf x^0,N,L)$ is by
saying that $n$ particles have a first passage time at the boundaries 
larger than $t$, while for $N-n$ ones the first passage time is smaller
than $t$. 
Thanks to Eq. (\ref{eq_surv_N}) we thus have
\begin{align}
p_n(t|\mathbf x^0,N,L)
=&\frac{1}{n!\;(N-n)!}
\sum_{\pi\in \Pi_N}
\int_t^\infty d t_1\cdots\int_t^\infty d t_n
\nonumber\\
&\int_0^t d t_{n+1}\cdots\int_0^t d t_N \prod_{k=1}^N\left[-\frac{d S_1(t_k|x_{\pi(k)}^0,1,L)}{d t_k}\right]
\nonumber\\
=&\frac{1}{n!\;(N-n)!}
\sum_{\pi\in \Pi_N}
\prod_{k=1}^n S_1(t|x_{\pi(k)}^0,1,L) \prod_{k=n+1}^N[1- S_1(t|x_{\pi(k)}^0,1,L)],
\label{eq_p_n}
\end{align}
where the combinatorial factors again arise as a consequence of the particles exchange symmetry.  
If the initial conditions $\mathbf x^0$ are randomly distributed
along the channel, all the single-particle survival probabilities in
Eq. (\ref{eq_p_n}) become equal and Eq. (\ref{eq_s_n}) neatly
simplifies into
\begin{equation}
\label{eq_s_n_simplified}
S_n(t|N,L)
=
\sum_{k=n}^N
\binom{N}{k}
\;S_1(t|1,L)^k
\;[1-S_1(t|1,L)]^{N-k},
\end{equation}
where 
\begin{equation}
S_1(t|1,L)\equiv
\frac{1}{L}\;\int_{-L/2}^{L/2}\;d x_{k}^0\;\;S_1(t|x_{k}^0,1,L)
\end{equation} 
is the survival probability of a free particle
initially placed at random within the channel.
To the best of our knowledge, we are accounting here the first
analytical derivation of Eq. (\ref{eq_s_n_simplified}), 
which governs the escape
dynamics of passive particles in Single File. 
In particular, the channel emptying
probability becomes
\begin{equation}
\label{eq_emptying}
1-S_1(t|N,L)
=
1-
\sum_{k=1}^N
\binom{N}{k}
\;S_1(t|1,L)^k
\;[1-S_1(t|1,L)]^{N-k}
\end{equation}
Integrating Eq. (\ref{eq_emptying}) over time gives the mean emptying time:
\begin{equation}
\label{eq_mean_emptying_time}
T_1(N,L)=\int_0^\infty d t\;S_1(t|N,L).
\end{equation}

\begin{figure}
\includegraphics[width=1\columnwidth]{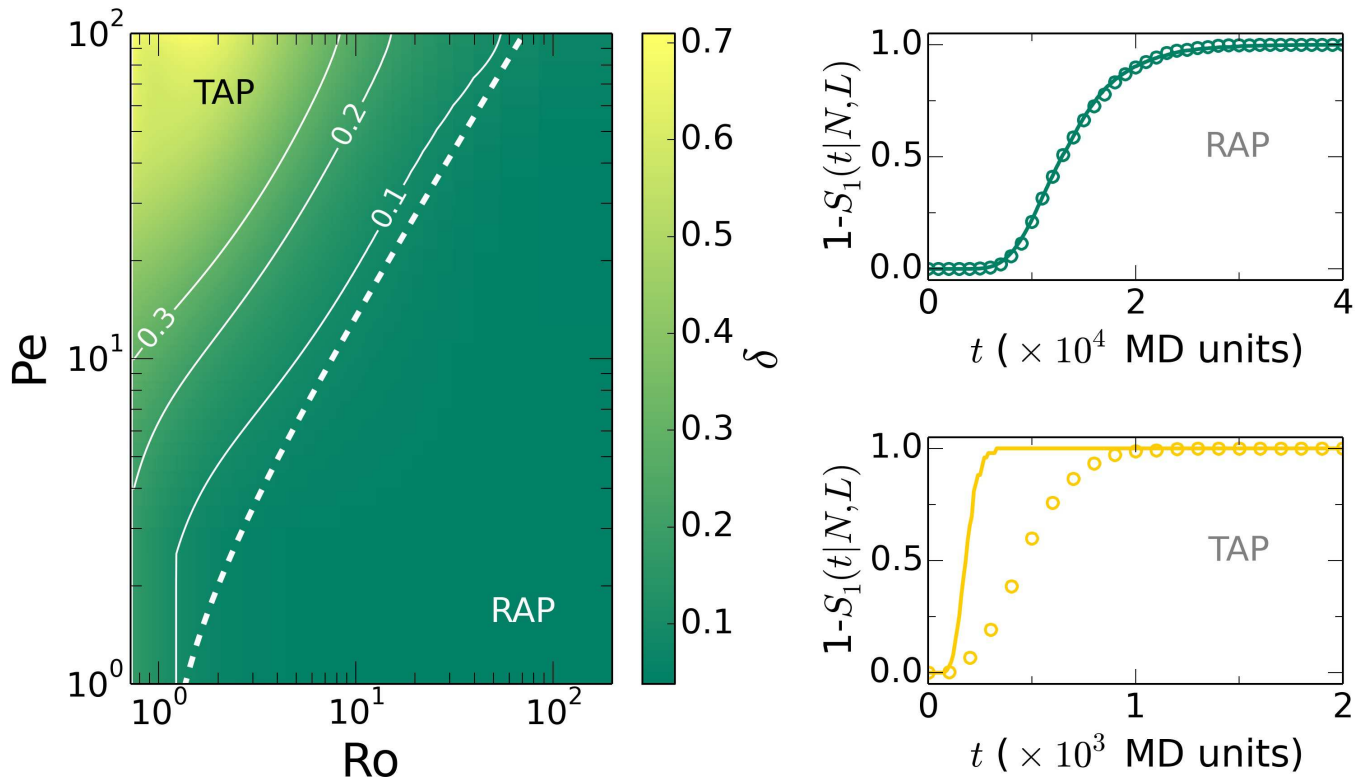}
\caption{
(Color online) 
Left panel: Ro-Pe phase diagram for the emptying process of an ASF system. Level curves refer to the relative difference $\delta(N,L)$ in Eq.~(\ref{eq:delta}) for $N=21$ and $L=52$ (MD units). 
The dashed curve, given by  Eq.~(\ref{eq:transition_line}), distinguishes
TAP's (above) from  RAP's (below). Right panels: Emptying probability $1-S_{1}(t|N,L)$ for a RAP ($\mbox{Ro}=200$, $\mbox{Pe}=50$) and a TAP ($\mbox{Ro}=0.75$, $\mbox{Pe}=50$). Circles are numerical simulations, lines are obtained through Eq.~(\ref{eq_s_n_simplified}).
}
\label{fig:fig_mean_time}
\end{figure}

\subsection{Unbiased channel emptying for active Brownian particles}
When activity is introduced, the interplay between rotational and
translational timescales has some important consequences. After a
collision two TAP's will tend to push one against the other, promoting
the formation of a small aggregate of swimmers.  Thus, collisions
among TAP's may be seen as non-reflecting  collisions; In turns, the 
the reflection principle 
could be effectively preserved in collisions among RAP's. 
As a consequence, 
Eqs.~(\ref{eq_s_n_simplified},\ref{eq_mean_emptying_time}) are
expected to play a role also in the realm of ASF systems, but only below the
crossover line established in Eq.~(\ref{eq:transition_line}).  

As specified in the Appendix, we performed simulations for different
values of Ro and Pe.  A quantitative way of addressing the
significance of
Eqs.~(\ref{eq_s_n_simplified},\ref{eq_mean_emptying_time}) is by
monitoring the the relative difference of the numerical mean emptying
time $T_{1}^{(num)}(N,L)$ with the theoretical result $T_{1}(N,L)$,
\begin{equation}
\delta(N,L) \equiv \frac{T_{1}^{(num)}(N,L) - T_{1}(N,L)}{T_{1}^{(num)}(N,L)},
\label{eq:delta}
\end{equation}
where the survival probability $S_1(t|1,L)$ in
Eq.~(\ref{eq_s_n_simplified}) has been numerically determined.  
The
level curves plots reported in Fig.~\ref{fig:fig_mean_time} show that,
indeed, below the line in Eq.~(\ref{eq:transition_line}) $\delta$ is
small and the agreement with
Eqs.~(\ref{eq_s_n_simplified},\ref{eq_mean_emptying_time}) is
remarkable for RAP's (See also top-right panel).  As we move deeper in
the TAP's region, $\delta$ grows of few percents and deviations with
Eqs.~(\ref{eq_s_n_simplified},\ref{eq_mean_emptying_time}) become
sensible (Compare also with the bottom-right panel).

\subsection{Dynamical clustering in Single File}

\begin{figure}
\includegraphics[width=1\columnwidth]{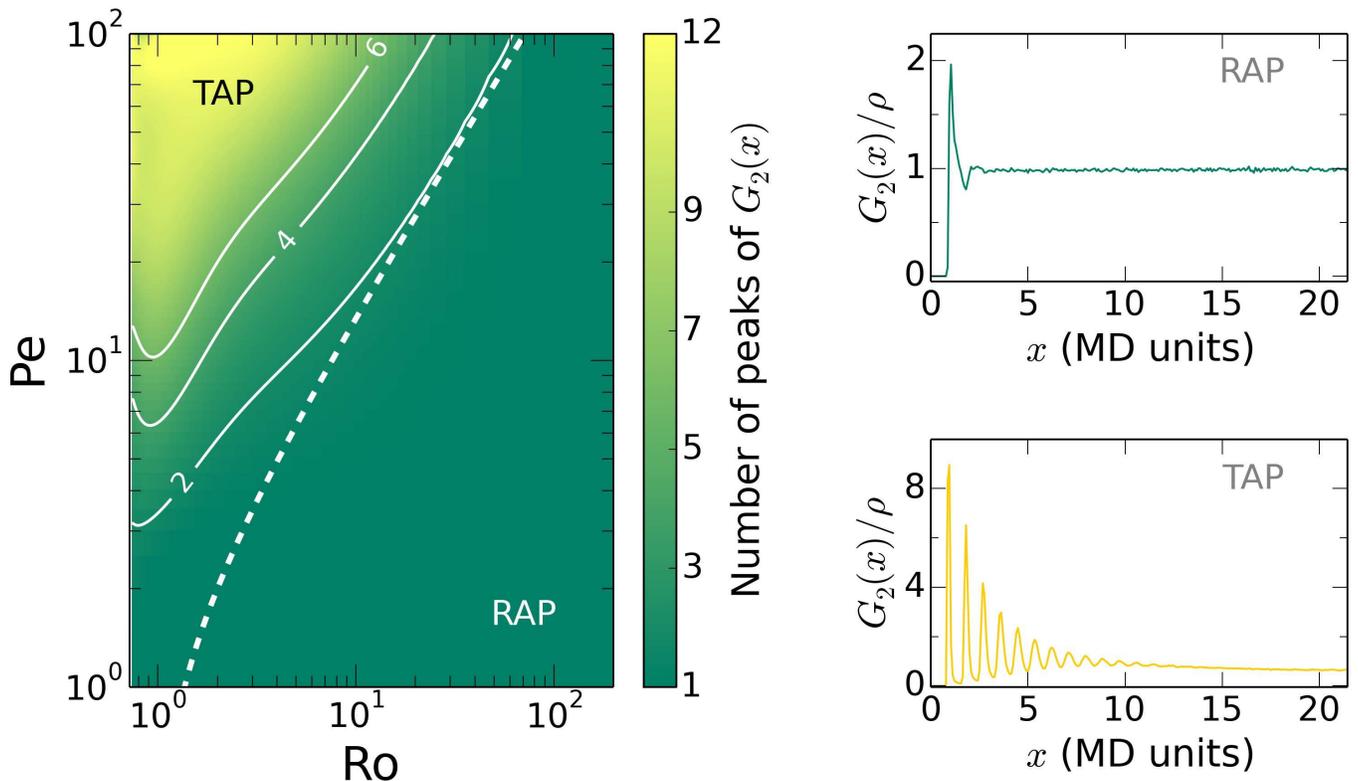}
\caption{(Color online) 
  Left panel: Phase diagram of the number of peaks of $G_{2}(x)$ 
  for ASF systems. $N=21$ hardcore particles of radius $R=0.5$ and   number density $\rho=N/L=0.5$, are simulated with periodic boundary conditions. 
The dashed curve, given by Eq.~(\ref{eq:transition_line}), distinguishes TAP's (above) from  RAP's (below). Full lines are level curves. 
  Right panels: radial distribution functions for
  RAP's ($\textrm{Ro}=200$, $\textrm{Pe}=50$) and TAP's ($\textrm{Ro}=0.75$, $\textrm{Pe}=50$).
  }
\label{fig:fig_radial}
\end{figure}

The validity-range of
Eqs.~(\ref{eq_s_n_simplified},\ref{eq_mean_emptying_time}) also corresponds
to a structural change in ASF
systems.  
Collisions can be understood as seeds for the
nucleation of dynamical clusters held together by the action of active
forces.  
Consider (active) hardcore particles of
radius $\sigma/2$. The presence of peaks  
at separation
multiple of $\sigma$ in the static, distinct radial
distribution function,
\begin{equation}
G_{2}(x) \equiv \frac{1}{N} 
\left\langle
\sum_{i \neq j} { \delta \left( x-x_{j}+x_{i} \right )  }
\right\rangle,
\label{eq:radial}
\end{equation} 
reveals the existence of clusters, with the number of peaks
corresponding to the typical cluster-size in units of $\sigma$.  We
numerically studied $G_2$ for an ASF system, this time with periodic
boundary conditions to increase our sampling.  Ensemble averages at a
time much larger than the system's correlation time are obtained by
considering the particles' coordinates $x_i$ and $x_j$ in different
dynamical realizations with random initial conditions.  Level curves
in Fig.~\ref{fig:fig_radial}, which report the number of peaks of
$G_2$ as a function of Ro and Pe, retain a close similarity with
Fig.~\ref{fig:fig_mean_time}. This means that RAP's do not aggregate,
whereas TAP's do. Hence, the line in Eq (\ref{eq:transition_line})
also marks the onset of dynamical clustering where SPP's continuously
merge and split into coherent assemblies.  Note that dynamical
clustering has been observed and discussed very recently in several
examples of self-propelled particle systems, in one, two and three
dimensions~\cite{Redner,Theurkauff,Palacci,Thompson, Mognetti_et_al_2013,Peruani,Cates_et_al_2014}.
In the following we show that this collective effect however does not improve the chances
of a single TAP to escape the channel against an external bias.

\begin{figure}
\centering
\includegraphics[width=.75\columnwidth]{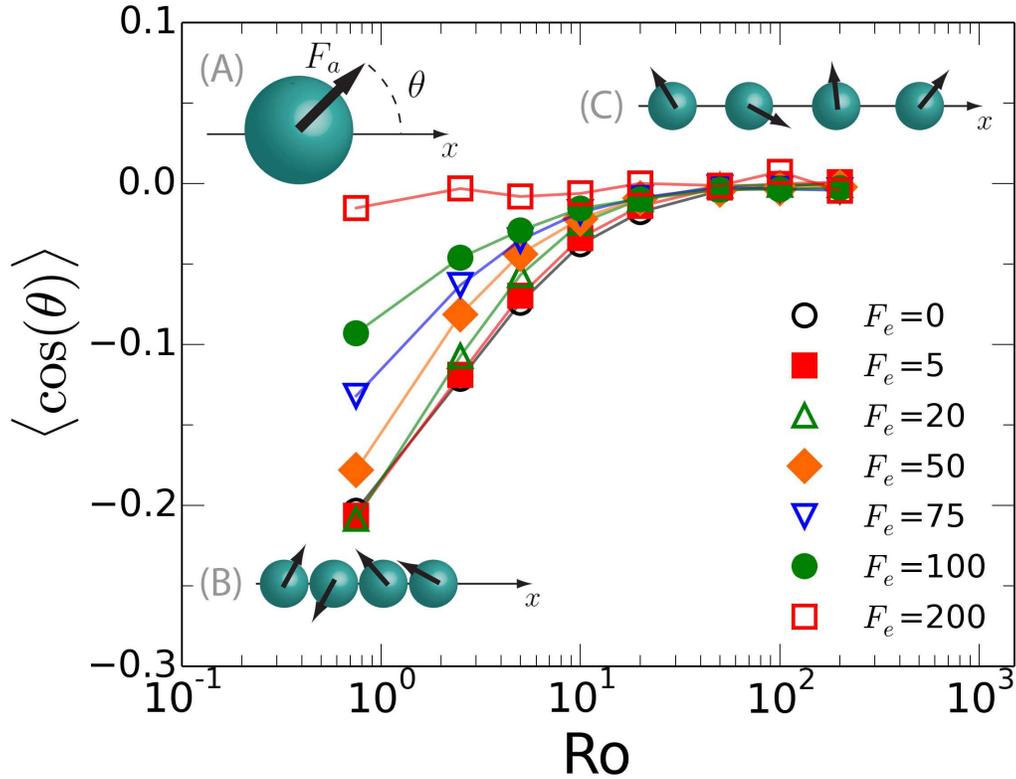}
\caption{ (Color online) Average direction of the clustered
  self-propelled particles at the instant in which a particle exits
  the channel from the left.  $F_e$ (in MD units) is the external
  force pointing to the right and $\theta$ is the angle between the
  active force orientation and the channel axis oriented to the right.
  The P\'eclet number is $\mbox{Pe} = 100$ and $N=21$.  (A) Sketch of
  the active particle model described in Section \ref{sec:model}. (B)
  TAP's leaving the channel against the bias $F_e$ form dynamically
  correlated clusters. (C) At variance, RAP's are uniformly dispersed
  with an active force with a randomly distributed orientation. (See
  animations in the Supplementary Material).}
\label{fig:fig_bias_angles}
\end{figure}

\subsection{Emptying process in the presence of an external bias}
An interesting practical 
situation is that of biased or ``tilted''
Single File channels. 
In several biological processes as well in microfluidic
devices SPP's may be found to move against the direction
of an externally imposed bias (e.g., the gravitational field~\cite{Wolff, Tailleur_2009}). 
We explore this problem
by adding to our ASF model an external constant and uniform force
field $F_{e}$ pointing along the channel direction, say, to its right
end.  Even within this extended parameter space
(enlarged by the presence of the field), the previous distinction
between TAP's and RAP's highlights 
different dynamical behaviors.  

We start the characterization of the emptying process for tilted channels
by studying the existence of possible correlations among the active force
orientations. 
Specifically, at the instant at which a particle leaves the
channel from the left (against the bias), we calculate the average $\langle\cos\theta\rangle$ 
($\theta$ being  the angle between the active force orientation and 
the channel axis) for those particle still in the channel that appear to be part of a cluster. 
We consider a particle to part of a cluster if one of its nearest neighbor is closer than 
the interaction range (See Appendix \ref{sec:simdetails}).  
The average is both over the different particles satisfying such a condition and over different 
escaping events. 
With this definition we can collect a good statistics for both TAP's and RAP's. 
In the latter case this happens because, although no stable cluster is formed,
within the channel there are enough colliding particles
which are forming an instantaneous two-particle cluster for the average to be significantly
calculated. 
In Fig.~\ref{fig:fig_bias_angles} we
show that at low Ro $\langle\cos\theta\rangle$ is smaller than zero, meaning
that at the instant when a particle leaves the channel against the bias
clustered TAP's in the channel tend to point along the
left. Increasing Ro towards RAP's particles, this tendency is lost. 
This observation is consistent with the fact that the 
``active clusters'' detected in the previous Section tend in fact to have a common orientation of
the active force at the moment a particle exits against the bias. 
Hence, particles in a cluster leave the channel one after the other in a rapid sequence
(See animations in the Supplementary Material).  
This effect, which also exists in the absence of the bias ($F_e=0$), 
gets weaker as $F_e$ increases, and  
vanishes for $F_e \gtrsim F_a$. 
Under the latter circumstances,   
trespassing the channel against the bias becomes a rare, isolated event.

\begin{figure}
\includegraphics[width=1.\columnwidth]{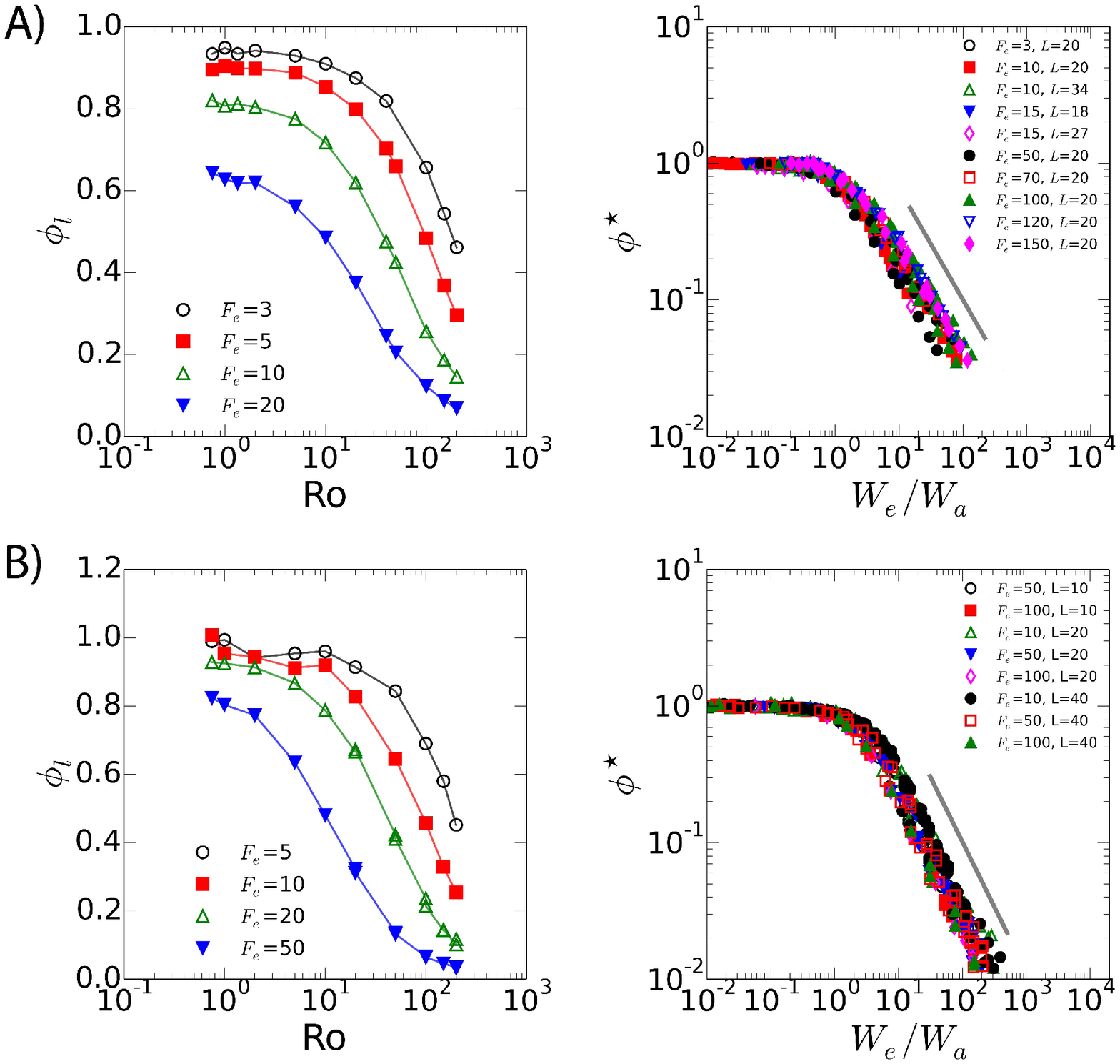}
\caption{(Color online) A) Left panel: capability of ASF particles to
  overcome external biases $F_e$ (in MD units).  The P\'eclet number
  is $\mbox{Pe} = 100$ and $N=21$.  Right panel: scaling symmetry
  characterizing $\phi^\star$ (See text for details). 
  The straight line highlights a power law decay with an
  exponent of about 0.75.  B) Same as A) with $N=1$. The straight
  line in the right panel has an exponent of about
  0.95.  }
\label{fig:fig_bias}
\end{figure}

We now focus on the capability of particles, initially randomly distributed
both in space and velocity direction,
to overcome the external bias. 
A quantitative measure of such a capability is given by 
(twice) the fraction of particles exiting on the left side of the
channel, 
\begin{equation}
\label{eq_fraction}
\phi_l({\rm Ro},{\rm Pe},F_e) \equiv2\;\frac{\langle N_l({\rm Ro},{\rm Pe},F_e)\rangle}{N},
\end{equation}
where $\langle N_l\rangle$ is the average number of particles absorbed
at the left end.  With the inclusion of the factor $2$ in
Eq.~(\ref{eq_fraction}), the value $\phi_l=1$ indicates that on
average half of the particles leave the channel from the left side and
the emptying process is not affected by $F_e$.  On the other hand,
when $\phi_l=0$ no particle is capable to overcome the bias.  The left
panel of Fig.~\ref{fig:fig_bias}a displays results of numerical
simulations performed for $21$ at $\textrm{Pe}=100$, a typical value
for several microswimmers.  Again, by increasing Ro we observe a
significant change of $\phi_l$.  Whereas TAP's retain a significant
capability of moving against the external bias even under Single File
conditions, RAP's are dominated by the force field.  Given the
similarity of the $\phi_l$ curves with respect to different biases in
the left panel of Fig.~\ref{fig:fig_bias}a, it is interesting to
check whether the SPPs exhibit a common way to trespass the channel
against a given $F_e$.  Intriguingly, by comparing the typical energy of
the SPP with the energy barrier imposed by the external bias, the
fraction of upstreamers $\phi_l$ turns out to satisfy a scaling
behavior whenever $F_a<F_e$. More precisely if, for given values of
${\rm Pe}$ and $F_e$, we divide $\phi_l$ by its maximum value
$\phi_{l}\:^{\rm max}({\rm Pe},F_e)=\phi_{l}({\rm Ro} \simeq 0,{\rm
  Pe},F_e)$, we obtain the fraction $\phi^{\star}_{l}$ of upstreamers
with respect to the fraction of TAP's (${\rm Ro} \simeq 0$),

\begin{equation}
\label{eq_rescaling}
\phi^\star_l({\rm Ro},{\rm Pe},F_e)\equiv
\frac{\phi_l({\rm Ro},{\rm Pe},F_e)}{\phi_{l}(0,{\rm Pe},F_e)}
=f\left(\frac{W_e}{W_a}\right),
\end{equation}

where
\begin{align}
\label{eq_works}
& W_e \equiv F_e ~ L/2,
&  \mbox{ and } &
& W_a({\rm Pe}, {\rm Ro}) \equiv k_B T \left[1+ \dfrac{{\rm Pe}^2}{{2\, \rm Ro}} \right] =k_B T ~\dfrac{D_a}{D_t}.
\end{align}
Notice that $W_e$ and $W_a$ can be understood  respectively as the average work done 
by the external force to bring a SPP out of a channel of length $L$, and the 
effective thermal energy for a SPP at given values of Pe and Ro. 
The right panel of Fig.~\ref{fig:fig_bias}a indeed exhibits a
data-collapse of simulations performed at different values of 
${\rm Ro}$, ${\rm Pe}$, $F_e$, $L$. 
The master curve $f$ in Eq.~(\ref{eq_rescaling}) shows an
initial very slow decay which crosses over to a power-law regime. 
In particular, $\phi^{\star}$ remains close to $1$ for TAP's (small ${\rm Ro}$ and thus large $W_a$). 
The existence of these two regimes demonstrate the 
specific capability of TAP's of being better resilient to the imposed field.
Moreover, the collapse of the data into a single master curve $f$ suggests that,
whenever a confined system fulfills Single File conditions, a common (universal) behavior for SPP's (either TAP's or RAP's) is expected regardless of their, dimensions 
or propelling mechanism. 
We believe this conclusion to be worth further inspected through experimental tests.

One may speculate whether the existence of clusters of TAP's improves the chances of trespassing the 
channel against the bias. Fig.~\ref{fig:fig_bias}b demonstrates that this is not the case. Essentially the
same behavior obtained for $N=21$ in  Fig.~\ref{fig:fig_bias}a is reproduced in Fig.~\ref{fig:fig_bias}b for $N=1$.
For the system sizes we have analyzed, this means that the capability of TAP's of moving against the bias is simply 
a consequence of their ``translational activity'' and is not a cooperative effect.

\section{Conclusion}
In this paper we have studied the dynamical properties of active
particles under Single File conditions, an ideal representation of a
very narrow channel in which particles cannot pass each other. Within
a simple model of self-propelled particles we have proposed a 
characterization of their dynamics by introducing,
besides the P\'eclet number Pe, the {\it rotation} number
Ro. 
Although instrumental to our scopes, Ro is simply related to other quantities 
(like the particles' persistence length) so that equivalent ways of characterizing
the translational-to-rotational character of the active particles are possible. 
According to our scheme, particles can be distinguished into TAP's and RAP's,
separated by Eq.~(\ref{eq:transition_line}) in the Ro-Pe phase space.
By examining data reported in the literature we find
that most bacteria and algae, even if characterized by a rotation
number Ro larger than that of spherical Janus particle, remain above the 
crossover line and behave as TAP's.  
The distinction between
RAP's and TAP's turns out to be important in investigating some
dynamical properties of active particles in Single File such as the
emptying process of a narrow channel with and without an external
bias.  In particular, we have shown that analytical tools here introduced to
characterize passive motion can be also used to describe
RAP's emptying process of the channel. Unlike RAP's the main
distinctive property of TAP's is their aptness to form aggregates,
named ``active clusters''. 
Although not in consequence of a cooperative effect, 
TAP's have been also shown to possess higher capabilities of trespassing 
narrow channels against external fields.  

Since our main goal has been to discuss within a simple picture how
classic Single File emptying processes are altered by the presence of activity,
throughout our study hydrodynamic interactions have been neglected.
These might be relevant especially in the case of bacteria and algae.
In particular, while for a single cell the hydrodynamic effect due to
extreme confinement could be neglected~\cite{Mannik},
cell-cell interactions in crowded systems may be influenced by
hydrodynamics~\cite{Drescher1}.  These last aspects are certainly
worth to be explored in the future, either experimentally or
theoretically.  
On the other hand, in cases in which physical interactions
could be dominated by steric rather than hydrodynamic
contributions, our findings are to be interpreted as a strategic
advantage for TAP's microorganisms in overcoming external biases in
strongly confined environment, leading to better chances in food
search, environment exploration or propagation of infections.  In this
view it is interesting to notice that genetic mutation may transform
TAP's into RAP's, thus, in principle, drug therapies increasing Ro by
enhancing the rotational diffusion could eliminate such an advantage.

\section*{Acknowledgments}
We gratefully acknowledge financial support from the European Research Council under the European Community's Seventh Framework Programme (FP7 2007-2013) / ERC Grant Agreement N. 297004 (DREOEMU), and from the University of Padova (PRAT 2013 N. CPDA138901).
In addition the authors kindly thank Prof. U. Keyser, Dr. S. Pagliara and Dr. P. Cicuta, from the University of Cambridge, for useful discussions.

\appendix

\section{Numerical simulations}
\label{sec:simdetails}
We briefly describe the numerical schemes used in our
simulations. 
We adopted natural Molecular Dynamics units for length, mass, and time, with $k_B=1$.
Between the collisions, we use a simple Brownian
Dynamics scheme:
\begin{equation}
  \left\{ \begin{array}{l}
    x(t+\Delta t) = x(t) + \dfrac{D_{t}}{k_{B}T} \left(F_a \cos(
    \theta(t)) + F_{e}\right)\Delta t 
    + \sqrt{\Delta t}\;\xi_t (t), 
    \\
    \\
    \theta(t+\Delta t) = \theta(t) + \sqrt{\Delta t}\;\xi_{r} (t),
  \end{array} \right.
\end{equation}
where $R=0.5$, $D_t=0.01$, $D_r$ is fixed by the specific value of Ro
considered, 
$T$=1 is the temperature of the
heath bath, $m$=1 is the particle mass and $\Delta t = 10^{-3}-
10^{-4}$ is the elementary time-step. 
The random noises $\xi_t (t)$
and $\xi_{r} (t)$ are normal independent pseudo-random numbers.
$F_a\in[1,300]$ and $F_e\in[0,300]$ are the constant active and external forces, respectively.
We performed simulations with both periodic and absorbing boundary
conditions. 
In the case of periodic boundaries the elapsing simulation time has
been $10^6 \Delta t$, while with absorbing boundaries 
it  has been variable 
in the range $[10^4\,\Delta t,10^7\, \Delta t]$; 
In both cases, from $10^3$ to $10^6$
independent realizations were performed. 

\begin{figure}[!h]
\centering
\includegraphics[width=0.5\columnwidth]{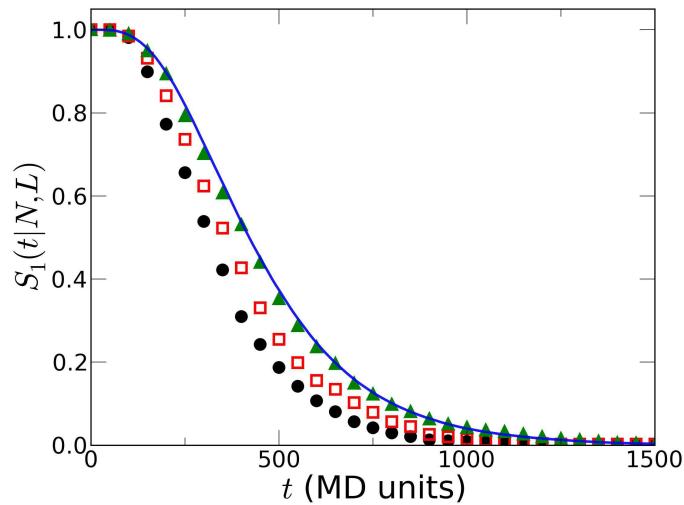}\\
\caption{(Color online) Emptying probability $S_1(t|N,L)$ obtained
using the two simulation methods described in the text, for {\it N}=5
particles, {\it L}=12 (MD units), Pe=50 and Ro=200. The full line is
the result of the point-like scheme, while the symbols are the result of
the truncated-shifted Lennard-Jones scheme. 
Different values of $\sigma$ are employed: full dots for
$\sigma$=1, open squares for $\sigma$=0.5, full triangles for
$\sigma$=0.1.  }
\label{fig:compare_sim}
\end{figure}

To simulate collisions between point-like particles, whenever
they overtake each other we switch
the labels of the colliding particles returning thus to their
pre-collision ordering. 
Conversely, to study the phenomenon of dynamical aggregation we simulate particles
with a certain size $\sigma$, employing a truncated-shifted
Lennard-Jones interaction potential \cite{Grest}:
\begin{equation}
 U_{ij} = 
\begin{cases}
4 \epsilon 
\left[ \left(\frac{\sigma}{r_{ij}} \right)^{12} -
  \left(\frac{\sigma}{r_{ij}} \right)^{6} \right] 
+ \epsilon & r_{ij}<\sigma \;2^{1/6}, \cr
0 & r_{ij} \geq \sigma \; 2^{1/6}. 
\end{cases}
\label{eq:LJ}
\end{equation}
Specifically, we have chosen $\sigma$=1. This two simulation methods
give the same results in the dilute limit $L/\sigma \to \infty$,
which can be achieved letting $\sigma \to 0$ with a fixed {\it
L}. Fig. \ref{fig:compare_sim} compares the emptying probabilities 
obtained using the two simulation methods.
Indeed, for $\sigma \to 0$ the second method gives the same result of
the first.

\end{document}